# Survival Strategies for African American Astronomers and Astrophysicists

JC Holbrook
*University of California, Los Angeles*

**Abstract**
The question of how to increase the number of women and minorities in astronomy has been approached from several directions in the United States including examination of admission policies, mentoring, and hiring practices. These point to departmental efforts to improve conditions for some of the students which has the overall benefit of improving conditions for all of the students. However, women and minority astronomers have managed to obtain doctorates even within the non-welcoming environment of certain astronomy and physics departments. I present here six strategies used by African American men and women to persevere if not thrive long enough to earn their doctorate. Embedded in this analysis is the idea of 'astronomy culture' and experiencing astronomy culture as a cross-cultural experience including elements of culture shock. These survival strategies are not exclusive to this small subpopulation but have been used by majority students, too.

## 1. African American History in Astronomy

Since 1955 only forty African Americans have earned doctorates in astronomy, astrophysics, or physics doing an astronomy dissertation (Fikes 2000; Williams and Oluseyi 2002; Valentine 2007). Considering the American Astronomical Society has 7000 members and the Bureau of Labor Statistics lists 1500 employed astronomers, depending upon which number you consider to be a true estimate of the number of astronomers and considering that three African American astronomers are deceased, then African Americans make up 2.47% or 0.53% of the PhDs (American Astronomical Society ; Bureau of Labor 2008). The statistics given by the Nelson Survey of the top 40 astronomy departments in the USA show that 1% of the faculty are African American which is 6 faculty out of 594 faculty for 2007 (Nelson 2007; Nelson 2007). Other minorities found in the Nelson Survey are 1.2% for Hispanic faculty which is 7 faculty, and 7.1% for Asian faculty which is 42 faculty. Increasing the number of women and minorities in astronomy has been a goal for more than thirty years. 94 women faculty out of 594 are part of the 2007 survey which is 15.8%.

The American Astronomical Society (AAS) has attempted to address this imbalance through the creation of a women's group (Committee on the Status of Women) and a minority group within AAS (Committee on the Status of Minorities 2002). Each of these has a newsletter that is freely available online that includes qualitative and quantitative studies about women and minorities in astronomy (Committee on the Status of Women 1999; Committee on the Status of 2001). These groups are meant to provide support for these minority astronomers and the newsletters are to share the latest research and opinions about diversifying astronomy.

Approaches to studying African American's success in the sciences has focused on the students as empty vessels which need to be filled with something that they are lacking, such as stronger math preparation, taking physics in secondary school, and completing a research project (Pearson Jr 1985; Pearson Jr and Pearson 1985; Pearson Jr and Bechtel 1989; Harding 1991; Hrabowski and Pearson Jr 1993; Pearson Jr and Fechter 1994; Harding 1998; Lewis and Collins 2001; Lewis, Pitts et al. 2002; Lewis 2003; Lewis and Connell 2005). My approach has been to consider African Americans success in astronomy as successfully navigating astronomy culture through framing it as a cross-cultural experience, moreover, as two cultures coming together with elements that conflict with each other. Disciplinary cultures in academia have been explored by social scientists for some decades and studies of the cultures of scientists in the United States and other parts of the world have increased with the creation of departments focused on the history and philosophy of science and science and technology studies

(Bourdieu 1990). Thus, the idea of an astronomy culture is consistent with these studies of other academic cultures. The issue with my cross-cultural approach is that African American culture is not monolithic: there are class, regional, occupational, and familial differences. Thus, the cultural conflicts experienced by individual African American students will differ from student to student.

**Culture Shock**

Culture Shock is defined as: "a sense of confusion and uncertainty sometimes with feelings of anxiety that may affect people exposed to an alien culture or environment without adequate preparation" (Dictionary.com 2011) and similarly as "a state of bewilderment and distress experienced by an individual who is suddenly exposed to a new, strange, or foreign social and cultural environment." (Dictionary.com 2011). The literature on culture shock focuses on cross-cultural encounters whereas I am considering it for intra-cultural encounters: the encounter with astronomy culture. Without examining specific events, individual agency, or external events, culture shock posits that an individual will go through a cycle of positive and negative feelings towards a new culture that is predictable. Culture shock was first formulated by the anthropologist Kalervo Oberg (Oberg 1954). He describes the four stages of culture shock as the Honeymoon phase, the Negotiation phase, the Adjustment phase, and the Mastery phase. The most anxiety and unhappiness occurs during the Negotiation and the Adjustment phases. The Negotiation phase has to do with learning to communicate, adjusting to new cultural norms, and as a result having feelings of loneliness and homesickness. The Adjustment phase concerns building new strategies and problem solving techniques to navigate the new culture as well as language acquisition: the new culture begins to make sense during this phase. Until the Mastery phase, the phase where the new culture is now normal, there is still anxiety that can be prolonged as long as the feeling of not understanding the culture remains.

**Astronomy Culture Insider and Outsider, Personal and General**

I was a student of physics and astronomy earning advanced degrees culminating in a PhD in astronomy and astrophysics in 1997. As a student assistant I worked on the Hubble Space Telescope, spent time at the Jet Propulsion Laboratory, and between my masters and my doctorate I worked at NASA Goddard Space Flight Center. These experiences define me as an astronomy culture 'insider', but since 1997 I have focused on the social and cultural aspects of astronomy as an anthropologist of astronomy, thus I am also an astronomy culture 'outsider' (Collins 1986; Harding 1991; Naples 1996). My research and activist activities keep me circulating in and through astronomy communities throughout the United States and internationally, allowing me to utilize and explore my insider and outsider status.

**Data Collection**

The data that I used to formulate the six strategies for survival was collected largely informally, gleaned from conversations and discussions that I had with astronomers over the last twenty years. For the International Year of Astronomy 2009, I produced a film 'Hubble's Diverse Universe' in which I interviewed eight other African American and Hispanic American astrophysicists (Holbrook and IYA 2009; Tapia 2009). In many of the interviews we discussed ideas and opinions about how to increase diversity within astronomy and we talked in detail about their own experiences in astronomy and as astronomers. I screened the film in physics and astronomy departments across the United States and after each screening there was discussion. Also, I had the opportunity to meet with students in a semi-confidential setting where we discussed issues including some of these survival strategies. As part of the American Physical Society's Gender Equity Conversation task force (Committee on the Status of Women in 2009), which focuses specifically on increasing the number of women in physics, the clash of cultures emerged as an element during the workshops that we conducted. Many of the elements paralleled those in astronomy. In June 2010 with my mentor's encouragement, I wrote down the six survival strategies because they were important for the interviews that we were conducting with women and minority astronomers for our current research project (Traweek 2009; Guillen 2010).

**The Six Survival Strategies' Broader Relevance**

Once formulated, the six survival strategies were sent to various African American astronomers and some women astronomers and I collected their feedback. The women astronomers commented that these strategies applied to them as well and to others that they knew. The African American astronomers provided details of which strategies that they used. Next, the six survival strategies were presented to a group of aspiring black South African students in Cape Town in Fall 2010. They provided feedback on which strategies they were already using to navigate astronomy culture. In summary, the six survival strategies have been validated by individual astronomers either as strategies that they use or that they have seen used by others. Therefore, the six survival strategies are relevant beyond the African American community extending to majority students, women, and in international contexts.

**1. Oblivious**

A person who tends to be oblivious to what is happening around them is one of the survival strategies. To be completely unaware of racist and sexist undertones and overtones is an effective strategy for surviving in a hostile environment. This could be an unintended strategy with a biological cause such as a form of Asperger syndrome, or it could be a behavior learned early on to ignore negative social interactions in favor of focusing exclusively on schoolwork.

As a person who travels through foreign cultures, I find that because I am American, I can be oblivious to the subtle racism found in other cultures. As a result, my life is better and less stressful when I am abroad. Such obliviousness is the result of being a stranger.

Another note, people who are aware of the sexism and racism often find their oblivious colleagues annoying because, well, they are oblivious. Especially annoying is when they insist that racism, sexism, and hostility do not exist. Instead it should be acknowledged that they could be employing a very successful survival strategy.

**2. Strong Familial Support**

A family that is supportive can greatly aid students who are experiencing hardships as astronomy students.  Family members can serve as a personal cheering section for the student reminding them that their family believe in them and believe that they can do astronomy, survive astronomy, and become an astronomer.  Family members can also provide financial support that may allow the student to study full time without the distraction of having to be a teaching assistant during the time they are preparing for their qualifying exams. However, most acknowledge that it is the emotional support rather than the financial support provided by their families that is the most important.

The astronomy culture found in the United States is such that family is considered to be a distraction from doing astronomy. Having a family, unless you are a man with a partner who is the primary caregiver, is perceived negatively. The connection that minority students have with their families is considered negatively as well. Minority students are encouraged to move away from their family. Indicating that within astronomy, the support provided by family is not acknowledged, nor is recognition given to the fact that such families produced the students who are now in astronomy therefore they must have been doing something right.

**3. Strong Departmental Support**

A mentor who believes that the student is capable of succeeding and helps them to succeed can help a student survive the most hostile environment.  Similar to familial support, the mentor becomes a cheering section for the student.  A mentor who is an astronomer who will actually say to other faculty in their department 'You cannot treat my student this way' is every minority student's dream.  Such a strong mentor who is willing to stand up to the rest of their department can transform a hostile environment by their actions, by forcing the hostile factions to not be blatant about their hostility. Ideally, students want the entire department to support them and believe in them, but just one person can compensate for this lack.

## 4. Divinely Inspired

In many ways religion is a taboo subject in astronomy. Some would argue that 'the creative design' debate has made this more so the case. Nonetheless, some minority students feel that they are studying astronomy because it is their calling, i.e. it is the reason that they were put on the earth and born at this moment in time and space. Doing astronomy or being an astronomer could be more or less divine in their mind, for example thinking that they have a gift for doing astronomy, to feeling or believing that the Gods want them to become an astronomer for some divine purpose. Some level of divine purpose gives students a resilience needed to survive being an astronomy student. Thus, when bad things happen such as racist encounters or simply facing challenging new material, those that are divinely inspired know that the racists cannot deter them from becoming an astronomer and know that they are destined to master the material.

## 5. Disconnect

A student may have the point of view that the environment that they are experiencing in graduate school or as college students is not the same as the environment that they will eventually be working in as an astronomer. The immediate culture may be so foreign, hostile, strange, etc, that they feel no connection and thus can consider themselves to be above the fray. 'These people are weird and I am not really one of them.' This is probably the equivalent of full blown culture shock, where the new culture is too strange, foreign, and repulsive to engage. In response, their education then becomes a transaction of goods and services rather than an acculturation process. They learn astronomy rather than how to be an astronomer. They are in survival mode where the process of getting a PhD becomes something to get through as quickly as possible. This disconnect allows them to distance themselves and thus be less impacted by a hostile environment. 'The haters cannot drive me out of astronomy.'

**6. Therapy and Medication**
Getting an advanced degree can be an extremely stressful experience, and there is sixth way to survive – therapy with or without chemical intervention. Asking professors today how graduate students in their generation dealt with the stress, many revealed that they drank a lot of alcohol – a form of self-medication. Many graduate students in all disciplines today end up on antidepressants. However, if being divinely inspired is a taboo discussion, then being on antidepressants falls into another magnitude of taboo. Nonetheless, therapy has provided a positive and safe venue for some African American astronomy students to discuss their struggles with someone who neither is judgmental nor disbelieving.

**Discussion**
I continue to discuss these survival strategies with women and minority astronomers and I am mindful of adding to the list, but thus far no new strategies have been suggested. Many admit to using a blending of many of these strategies rather than relying on just one or two. On the other hand, these strategies are born in response to conflict or hostility and thankfully not every African American astronomer has had that experience.

Going back to having a supportive departmental mentor being what every student really wants, there is the expectation that this is the right of every student in astronomy. As professors, mentoring is part of their job description. But, as one African American astrophysicist said, "Before meeting [my mentor], before sort of joining his department, the mentors, well, they were called mentors," implying that all mentors are not created equal. Isolation is a common experience among minority astronomy students and a dedicated mentor can do a lot to remove that sense of isolation. Isolation is often the first indication that a student is unhappy and considering dropping out of astronomy.

If the four phases of culture shock are layered onto the experiences of African American astronomy students, then the expectation is that they will be the most likely to consider abandoning astronomy during the Negotiation and Adjustment phases as they are struggling to learn the language, the norms, and how to navigate astronomy culture. The survival strategy of disconnect seems to arise during these two phases.

**Conclusions**

Survival strategies among women and minority astronomy students are varied but all can lead to successfully navigating what can be a very hostile and unsupportive environment. Thus, I am careful not to rank these strategies or place value judgments on them, for example considering which strategies are more or less healthy. However, I think it is fruitful to discuss these strategies in astronomy communities especially if departments are serious about being more inclusive, not just inclusive to African Americans but to all non-majority students.

I see two things that can be considered by astronomers that could move astronomy culture towards being more inclusive. First, astronomers need to acknowledge that the experience of individual students is not the same, and that this is especially true for women and minority students. What can be a healthy supportive environment for one student can be hostile and detrimental to another student. Acknowledging this difference is important and can lead to focusing on the individual needs of students allowing more to succeed. Second, returning to the survival strategies, astronomers can lend support to some of these strategies and change their departmental culture to make sure that other of these strategies are less often utilized. For example, professors should be on the lookout for non-majority students who are isolating themselves and be ready to offer support and encouragement at this crucial point.

There is no simple solution or band-aid that will suddenly increase the number of women and minority astronomers and diversify astronomy; however, analyzing astronomy culture and individual experiences may catalyze a cultural change that may bring such a goal within reach.

American Astronomical Society. "American Astronomical Society." from [www.aas.org](www.aas.org).
Bourdieu, P. (1990). <u>Homo Academicus</u>. Palo Alto, Stanford University Press.
Bureau of Labor, S. (2008). "Astronomers Wages and Density USA." from [http://www.bls.gov/oes/current/oes192011.htm](http://www.bls.gov/oes/current/oes192011.htm).
Collins, P. H. (1986). "Learning from the Outsider Within: The Sociological Significance of Black Feminist Thought." <u>Social problems</u> **33**(6): S14-S32-S14-S32.
Committee on the Status of, M. (2001). "Aas Committee on the Status of Minorities." from [http://csma.aas.org/panchromatic.html](http://csma.aas.org/panchromatic.html).
Committee on the Status of Minorities. (2002). "Spectrum: Newsletter of the Committee on the Status of Minorities in Astronomy." from [http://csma.aas.org/spectrum.html](http://csma.aas.org/spectrum.html).
Committee on the Status of Women. "Aas Committee on the Status of Women." Retrieved November 24, 2010, 2010, from [http://www.aas.org/cswa/](http://www.aas.org/cswa/).
Committee on the Status of Women. (1999). "Aaswomen Newsletter." from [http://www.aas.org/cswa/AASWOMEN.html](http://www.aas.org/cswa/AASWOMEN.html).
Committee on the Status of Women in, P. (2009). "Conversations on Gender Equity Site Visits." from [http://www.aps.org/programs/women/workshops/gender-equity/sitevisits/index.cfm](http://www.aps.org/programs/women/workshops/gender-equity/sitevisits/index.cfm).
Dictionary.com (2011). Culture Shock. <u>Dictionary.com Unabridged</u>, Random House, Ind.
Dictionary.com (2011). Culture Shock. <u>Merriam-Webster's Medical Dictionary</u>.
Fikes, R. (2000). "Careers of African Americans in Academic Astronomy." <u>The Journal of Blacks in Higher Education</u>(29): 132-134.
Guillen, R. (2010). "Welcome to the Frontpage." from [http://nsf.teknoculture.com/](http://nsf.teknoculture.com/).
Harding, S. G. (1991). <u>Whose Science? Whose Knowledge?: Thinking from Women's Lives</u>, Cornell Univ Pr.
Harding, S. G. (1998). <u>Is Science Multicultural?: Postcolonialisms, Feminisms, and Epistemologies</u>, Indiana Univ Pr.
Holbrook, J. and IYA (2009). <u>Iya2009usa: Cultural Astronomy and Storytelling Working Group</u>. Bulletin of the American Astronomical Society.
Hrabowski, F. A. and W. Pearson Jr (1993). "Recruiting and Retaining Talented African-American Males in College Science and Engineering." <u>Journal of College Science Teaching</u> **22**(4): 234-238.
Lewis, B. F. (2003). "A Critique of Literature on the Underrepresentation of African Americans in Science: Directions for Future Research." <u>Journal of Women and Minorities in Science and Engineering</u> **9**(3/4).

Lewis, B. F. and A. Collins (2001). "Interpretive Investigation of the Science-Related Career Decisions of Three African-American College Students." Journal of research in science teaching **38**(5): 599-621.

Lewis, B. F. and S. Connell (2005). "African American Students' Career Considerations and Reasons for Enrolling in Advanced Science Courses." Negro Educational Review, The **56**(2-3): 221-231.

Lewis, B. F., V. R. Pitts, et al. (2002). "A Descriptive Study of Pre-Service Teachers'perceptions of African-American Students'ability to Achieve in Mathematics and Science." Negro Educational Review **53**(1/2): 31-42.

Naples, N. A. (1996). "A Feminist Revisiting of the Insider/Outsider Debate: The "Outsider Phenomenon" In Rural Iowa." Qualitative Sociology **19**(1): 83-106.

Nelson, D. (2007). "Astronomy Faculty." 2011, from http://chem.ou.edu/~djn/diversity/astrodiv.html.

Nelson, D. (2007). "Astronphd.Html." 2011, from http://chem.ou.edu/~djn/diversity/PhDTables/01astronPhD.html.

Oberg, K. (1954). Culture Shock, Bobbs-Merrill.

Pearson Jr, W. (1985). Black Scientists, White Society, and Colorless Science, New York: Associated Faculty Press.

Pearson Jr, W. and H. K. Bechtel (1989). Blacks, Science, and American Education, Rutgers University Press, 109 Church Street, New Brunswick, NJ 08901 ($35.00).

Pearson Jr, W. and A. Fechter (1994). Who Will Do Science? Educating the Next Generation, Johns Hopkins University Press, 2715 North Charles Street, Baltimore, MD 21218-4319.

Pearson Jr, W. and L. R. C. Pearson (1985). "Baccalaureate Origins of Black American Scientists: A Cohort Analysis." Journal of Negro Education: 24-34.

Tapia, D. (2009). Hubble's Diverse Universe. USA**:** 38 minutes.

Traweek, S. (2009). "Award#0956589 - Women and Minority Astronomers Strategic Engagement with Distributed, Multi-Disciplinary Collaborations and Large Scale Databases." from http://www.nsf.gov/awardsearch/showAward.do?AwardNumber=0956589.

Valentine, J. (2007). "Black Women in Physics." from http://www.pha.jhu.edu/%7Ejami/bwip.html.

Williams, S. and H. Oluseyi. (2002). "Who Are the Black Astronomers and Astrophysicists?", from http://www.math.buffalo.edu/mad/physics/astronomy-peeps.html.